# Mechanical coupling effects of 2D lattices uncovered by decoupled micropolar elasticity tensor and symmetry operation

Zhiming Cui and Jaehyung Ju

UM-SJTU Joint Institute, Shanghai Jiao Tong University

800 Dongchuan Road, Shanghai 200240, China

**Abstract**

Mechanical couplings such as axial–shear and axial–bending have great potential in the design of active mechanical metamaterials with directional control of input and output loads in sensors and actuators. However, the current ad hoc design of mechanical coupling without theoretical support of elasticity cannot provide design guidelines for mechanical coupling with lattice geometries. Moreover, the correlation between mechanical coupling effects and geometric symmetry is not yet clearly understood. In this work, we systematically search for all possible mechanical couplings in 2D lattice structures by determining the non-zero diagonal terms in the decomposed micropolar elasticity tensor. We also correlate the mechanical couplings with the point-group symmetry of 2D lattices by applying the symmetry operation to the decomposed micropolar elasticity tensor. The decoupled micropolar constitutive equation uncovers eight coupling effects for 2D lattice structures. The symmetry operation of the decoupled micropolar elasticity tensor reveals the correlation of the mechanical coupling with the point groups. Our findings can strengthen the design of mechanical metamaterials with potential applications in areas including sensors, actuators, soft robots, and active metamaterials for elastic/acoustic wave guidance and thermal management.

**Keywords**: mechanical coupling, point groups, symmetry operation, micropolar elasticity, mechanical metamaterials

1. Introduction

Poisson's ratio [1] is a well-known mechanical coupling effect describing the coupling of longitudinal and lateral strain. In the earlier stage, Poisson and Navier believed that Poisson's ratio was 0.25 for all isotropic materials based on the uni-constant theory [2]. Later, the theoretical bound of Poisson's ratio was determined to be −1 to 0.5 for 3D isotropic materials using the positive-definite property of the elasticity tensor [3,4]. Since Lakes' experimental demonstration of negative Poisson's ratio of cellular foams in 1987 [5], numerous researchers have explored the engineering of the mechanical coupling with lattice (or cellular) structures in stochastic or non-stochastic manners [2,6–12]. In addition to Poisson's ratio, another mechanical coupling that can be described by the constitutive equation of the classical Cauchy elasticity is the axial–shear coupling [13–15], describing a shear deformation coupled with an axial input loading.

Recently, several mechanical coupling effects have been reported: axial–twist [16], axial–rotation [17], and axial–bending [18] couplings. These unique coupling effects have potential applications in mechanical cloaking [19–21], sensors/actuators [22], deployable structures [23], and elastic/acoustic wave control [17,24–26]. These couplings cannot be explained by the classical Cauchy elasticity, which uses the symmetric stress (or strain) tensor. Instead, the micropolar elasticity with an asymmetric stress tensor [27,28] provides more freedom to search mechanical coupling effects by adapting the independent rotation of vertices.

The coupling effects are related to the non-zero terms of the micropolar elasticity tensor [15,17]. Liu et al. proposed a planar isotropic micropolar elasticity tensor to capture the 2D chiral effect, introducing a material parameter related to chirality that represents the axial–rotation coupling [17]. From the orthotropic micropolar elasticity model proposed in [27], Yuan et al. found another parameter related to axial–shear coupling and proposed an equation to quantitatively measure the magnitude of the coupling effect [15]. However, these ad hoc design-based micropolar mechanics models adapt a mixed shear strain and nodal rotation, requiring identification of all possible mechanical couplings in the generalized lattice domain.

From observations of previous coupling effects [15–18], chirality, the broken mirror symmetry is vital in generating mechanical coupling effects. A triangular chiral [3] or hexa-chiral lattice composed of circular rings and straight ligaments has an axial–rotation coupling. Notably, chirality refers to lattices with broken mirror symmetries [29,30], which is a significant factor causing an axial–shear coupling. Yuan et al. showed that an axial–shear coupling exists in a tetra-chiral lattice but not in a tetra-achiral lattice [15]. Li et al. designed a square chiral lattice by assigning two materials to the structs while breaking a mirror symmetry, producing an axial–shear coupling [14]. Cui et al. showed that the non-centrosymmetric square lattice has axial–bending coupling [18]. However, there is no clear understanding of the correlation between mechanical coupling effects and geometric symmetry.

Therefore, this work aims to systematically search all possible mechanical couplings in 2D lattice structures. We construct a micropolar elasticity tensor that can decouple the shear strain and nodal rotation, explicitly identifying the coupling effects in the constitutive equation by obtaining non-zero diagonal terms in the decomposed micropolar elasticity tensor. Following Neumann's principle [31], we obtain the micropolar elasticity tensor for varying symmetry operations, providing possible coupling effects for varying point groups of lattices. We uncover all possible mechanical coupling effects of 2D lattices, which are strongly supported by the micropolar elasticity and subsequently validated by finite element (**FE**)-based simulations.

## 2. Decoupled micropolar elasticity tensor

The constitutive equation of the micropolar elasticity can be written as

$$\boldsymbol{\sigma} = \mathbf{C}:\boldsymbol{\varepsilon} + \mathbf{B}:\boldsymbol{\kappa} \ (\sigma_{ij} = C_{ijkl}\varepsilon_{kl} + B_{ijkl}\kappa_{kl})$$

$$\mathbf{m} = \mathbf{B}:\boldsymbol{\varepsilon} + \mathbf{D}:\boldsymbol{\kappa} \ (m_{ij} = B_{klij}\varepsilon_{kl} + D_{ijkl}\kappa_{kl}), \quad (1)$$

where $\varepsilon_{ij} = \frac{\partial u_j}{\partial x_i} - e_{kij}\varphi_k$ is the Cauchy strain tensor and $\kappa_{ij} = \frac{\partial \varphi_j}{\partial x_i}$ is the couple strain tensor, i.e., the bending curvature [32]. Correspondingly, $\sigma_{ij}$ is the Cauchy stress tensor, and $m_{ij}$ is the couple stress tensor. Note that $\varepsilon_{ij}$ and $\sigma_{ij}$ are no longer symmetric. In addition, note that **B** contribute to both the Cauchy stress by nodal rotation $\boldsymbol{\kappa}$ and the couple stress by the Cauchy strain $\boldsymbol{\varepsilon}$.

Assembling the Cauchy and couple stress tensors as a vector form in planar motion [28], we can express the 2D micropolar constitutive equation as

$$\boldsymbol{\sigma} = \mathbf{Q} \cdot \boldsymbol{\varepsilon}$$

$$\begin{Bmatrix} \sigma_{11} \\ \sigma_{22} \\ \sigma_{12} \\ \sigma_{21} \\ m_{13} \\ m_{23} \end{Bmatrix} = \begin{bmatrix} C_{11} & C_{12} & C_{13} & C_{14} & B_{11} & B_{12} \\ & C_{22} & C_{23} & C_{24} & B_{21} & B_{22} \\ & & C_{33} & C_{34} & B_{31} & B_{32} \\ & & & C_{44} & B_{41} & B_{42} \\ & sym & & & D_{11} & D_{12} \\ & & & & & D_{22} \end{bmatrix} \begin{Bmatrix} \varepsilon_{11} \\ \varepsilon_{22} \\ \varepsilon_{12} \\ \varepsilon_{21} \\ \kappa_{13} \\ \kappa_{23} \end{Bmatrix}, \quad (2)$$

where $\mathbf{Q}$ represents the 2D micropolar elasticity tensor. We combine $\sigma_{ij}$ and $m_{ij}$ for a generalized stress vector form; $\boldsymbol{\sigma} = \{\sigma_{11}, \sigma_{22}, \sigma_{12}, \sigma_{21}, m_{13}, m_{23}\}^T = \{\sigma_1, \sigma_2, \sigma_3, \sigma_4, \sigma_5, \sigma_6\}^T$. Similarly, we can combine $\varepsilon_{ij}$ and $\kappa_{i3}$ for a generalized strain vector form given by $\boldsymbol{\varepsilon} = \{\varepsilon_{11}, \varepsilon_{22}, \varepsilon_{12}, \varepsilon_{21}, \kappa_{13}, \kappa_{23}\}^T = \{\varepsilon_1, \varepsilon_2, \varepsilon_3, \varepsilon_4, \varepsilon_5, \varepsilon_6\}^T$.

The non-symmetric tensors $\sigma_{ij}$ and $\varepsilon_{ij}$ in the micropolar theory can still be decomposed into symmetric and anti-symmetric parts [33,34], which are

$$S_{ij} = \frac{\sigma_{ij} + \sigma_{ji}}{2}, \quad T_{ij} = \frac{\sigma_{ij} - \sigma_{ji}}{2} \quad (3)$$

$$E_{ij} = \frac{\varepsilon_{ij} + \varepsilon_{ji}}{2}, \quad A_{ij} = \frac{\varepsilon_{ij} - \varepsilon_{ji}}{2}. \quad (4)$$

Apparently, $S_{12} = S_{21}$, $T_{12} = -T_{21}$, and $E_{12} = E_{21}$, $A_{12} = -A_{21}$. Notably, there are still six independent variables for each decoupled stress and strain tensor: $\{S_{11}, S_{22}, S_{12}, T_{12}, m_{13}, m_{23}\}^T$ and $\{E_{11}, E_{22}, E_{12}, A_{12}, \kappa_{13}, \kappa_{23}\}^T$. The physical meaning of $T_{12}$ is the torque per unit area directly applied on the nodes, which is usually zero in typical loading conditions, e.g., axial and shear loadings.

It is worthwhile to mention that

$$E_{12} = E_{21} = \frac{\varepsilon_{12} + \varepsilon_{21}}{2} = \frac{u_{1,2} + u_{2,1}}{2} = \frac{\gamma_{12}}{2}$$

$$A_{12} = -A_{21} = \frac{\varepsilon_{12} - \varepsilon_{21}}{2} = \frac{u_{2,1} - u_{1,2}}{2} - \varphi = \Omega - \varphi, \quad (5)$$

where $\Omega$ is the macro-rotation induced by the displacement field. The physical meaning of $E_{12}$ is half of the engineering shear strain, whereas $A_{12}$ represents a local rotation at one point.

Note that the macroscopic rotation $\Omega$ is different from the local rotation $A_{21}$. Considering the deformation of a single beam, as shown in Figure 1, the macro-rotation $\Omega$ is explicitly dependent on the displacements of the join5. However, the local rotation $A_{21}$ is inexplicitly affected by the displacements and is independent of $\Omega$. The nodal rotation $\varphi$ is the sum of $\Omega$ and $A_{21}$, which is the net rotation at the joints.

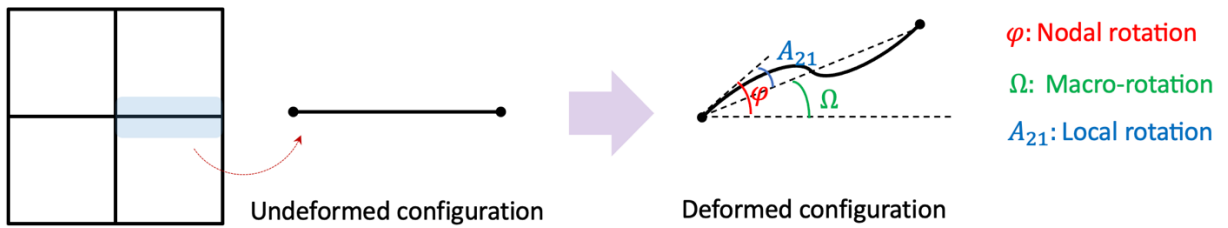

Figure 1. Physical meanings of local rotation $A_{12}$, macro-rotation $\Omega$, and nodal rotation $\varphi$ ($= \Omega + A_{21}$).

We can rearrange Equation (2) into a constitutive equation with decoupled stress and strain components:

$$\mathbf{S} = \mathbf{q} \cdot \mathbf{E}$$

$$\begin{Bmatrix} S_{11} \\ S_{22} \\ S_{12} \\ T_{12} \\ m_{13} \\ m_{23} \end{Bmatrix} = \begin{bmatrix} c_{11} & c_{12} & c_{13} & c_{14} & b_{11} & b_{12} \\ & c_{22} & c_{23} & c_{24} & b_{21} & b_{22} \\ & & c_{33} & c_{34} & b_{31} & b_{32} \\ & & & c_{44} & b_{41} & b_{42} \\ & sym & & & d_{11} & d_{12} \\ & & & & & d_{22} \end{bmatrix} \begin{Bmatrix} E_{11} \\ E_{22} \\ 2E_{12} \\ 2A_{12} \\ \kappa_{13} \\ \kappa_{23} \end{Bmatrix}. \tag{6}$$

We denote the decoupled elasticity tensor **q** whose individual components are represented by lower case letters. Note that we use $2E_{12}$ and $2A_{12}$ in the strain vector rather than $E_{12}$ and $A_{12}$ to make **q** symmetric. Also note that the submatrix composed of the first three rows and columns of **q** is equivalent to the classical Cauchy elasticity tensor.

We can correlate the two elasticity tensors **q** and **Q** using Equations (3) and (4):

$$\begin{bmatrix} c_{11} & c_{12} & c_{13} & c_{14} & b_{11} & b_{12} \\ & c_{22} & c_{23} & c_{24} & b_{21} & b_{22} \\ & & c_{33} & c_{34} & b_{31} & b_{32} \\ & & & c_{44} & b_{41} & b_{42} \\ & sym & & & d_{11} & d_{12} \\ & & & & & d_{22} \end{bmatrix} =$$

$$\begin{bmatrix} C_{11} & C_{12} & \frac{C_{13}+C_{14}}{2} & \frac{C_{13}-C_{14}}{2} & B_{11} & B_{12} \\ & C_{22} & \frac{C_{23}+C_{24}}{2} & \frac{C_{23}-C_{24}}{2} & B_{21} & B_{22} \\ & & \frac{C_{33}+C_{44}+2C_{34}}{4} & \frac{C_{33}-C_{44}}{4} & \frac{B_{31}+B_{41}}{2} & \frac{B_{32}+B_{42}}{2} \\ & & & \frac{C_{33}+C_{44}-2C_{34}}{4} & \frac{B_{31}-B_{41}}{2} & \frac{B_{32}-B_{42}}{2} \\ & sym & & & D_{11} & D_{12} \\ & & & & & D_{22} \end{bmatrix}. \tag{7}$$

By decoupling the shear strain $E_{12}$ and local rotation $A_{12}$, we obtain a decoupled micropolar elasticity tensor, revealing a clear physical meaning for each term, as shown in Figure 2. Figure 2 shows the decoupled micropolar elasticity tensor uncovering mechanical coupling effects. We find that the coupling effects are related to the non-diagonal terms of the decoupled elasticity tensor. There are eight mechanical coupling effects for 2D lattice materials: an axial–axial (A–A) coupling, which is simply the Poisson effect; an axial–shear (A–S) coupling; an axial–rotation (A–R) coupling; an axial–bending (A–B) coupling; a shear–rotation (S–R) coupling; a shear–bending (S–B) coupling; a bending–rotation (B–R) coupling; and a bending–bending (B–B) coupling. The A–A, A–S, A–R, and A–B coupling have been previously discovered [13,15,17,18]. However, the remaining four coupling effects, S–R, S–B, B–R, and B–B, have not yet been explored in 2D lattice structures.

Notably, the diagonal terms of the decoupled micropolar elasticity tensor are associated with a direct effect on the resistance to deformation: the axial moduli ($c_{11}$ and $c_{22}$), shear modulus ($c_{33}$), and bending stiffness ($d_{11}$ and $d_{22}$). However, the non-diagonal terms represent the mechanical coupling effects. The term $c_{12}$ controls Poisson's effect. The terms $c_{13}$ and $c_{23}$ affect the A–S coupling: e.g., a shear deformation by a uniaxial loading. The terms $c_{14}$ and $c_{24}$ affect the A–R coupling, where the rotation refers to the nodal rotation $A_{12}$. The nonsymmetric matrix $\begin{bmatrix} b_{11} & b_{12} \\ b_{21} & b_{22} \end{bmatrix}$ indicates an A–B coupling. The term $d_{12}$ contributes to a B–B coupling, i.e., bending in one direction led by bending in another direction.

The terms $b_{41}$ and $b_{42}$ represent a B–R coupling. The term $c_{34}$ guides an S–R coupling. The terms $b_{31}$ and $b_{32}$ determine an S–B coupling.

$$\begin{Bmatrix} S_{11} \\ S_{22} \\ S_{12} \\ T_{12} \\ M_{13} \\ M_{23} \end{Bmatrix} = \begin{bmatrix} c_{11} & c_{12} & c_{13} & c_{14} & b_{11} & b_{12} \\ & c_{22} & c_{23} & c_{24} & b_{21} & b_{22} \\ & & c_{33} & c_{34} & b_{31} & b_{32} \\ & & & c_{44} & b_{41} & b_{42} \\ & sym & & & d_{11} & d_{12} \\ & & & & & d_{22} \end{bmatrix} \begin{Bmatrix} E_{11} \\ E_{22} \\ 2E_{12} \\ 2A_{12} \\ \kappa_{13} \\ \kappa_{23} \end{Bmatrix}$$

- Axial-Axial (A-A) coupling
- Axial-Shear (A-S) coupling
- Axial-Rotation (A-R) coupling
- Axial-Bending (A-B) coupling
- Shear-Rotation (S-R) coupling
- Shear-Bending (S-B) coupling
- Bending-Rotation (B-R) coupling
- Bending-Bending (B-B) coupling

*Figure 2. Constitutive relation with the decoupled micropolar elasticity tensor to show mechanical coupling effects*

Figure 3 shows the eight potential mechanical coupling effects for 2D structures. Notably, the constitutive relation with the Cauchy elasticity tensor can still identify the A–A and A–S couplings [35] yet has a limitation in the search for other coupling effects. In contrast, the micropolar elasticity can recognize the rest of the couplings. The A–S coupling has been found in the tetra-chiral structures [14,15,36]. A triangular chiral structure is known to have an A–R coupling [17]. Recently, an A–B coupling was observed in a non-centrosymmetric square lattice [18]. Note that the A–B and S–B couplings in this work are different from the ones in the laminated composites [37]; the bending in this study is in the in-plane deformation, whereas the bending in the laminated composites is in the out-of-plane deformation.

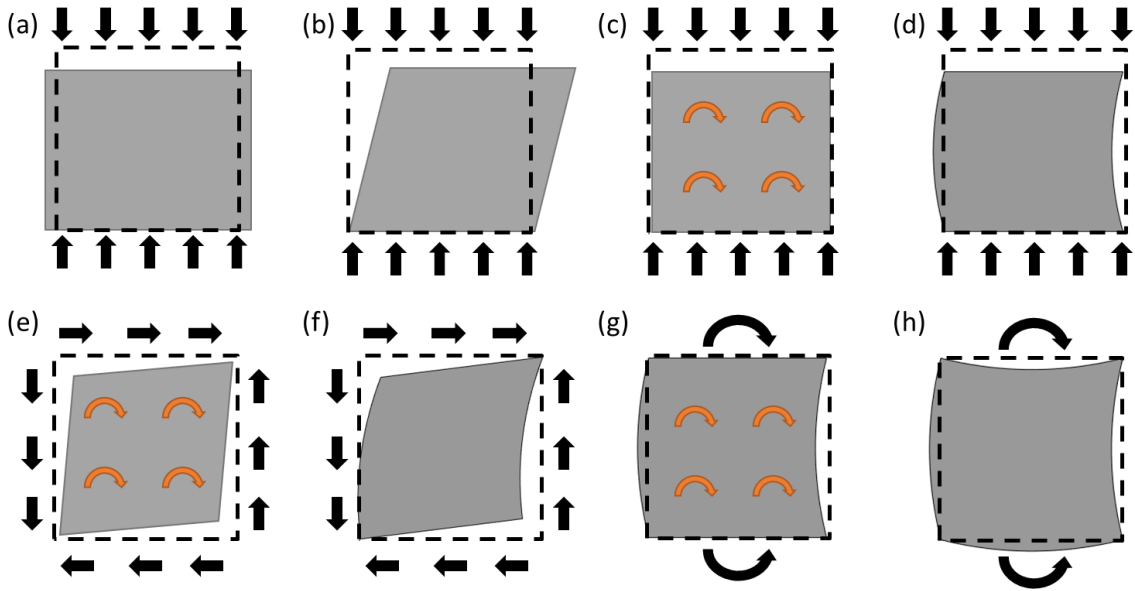

*Figure 3. Visual interpretation of the 2D mechanical coupling effects: (a) axial–axial (A–A) coupling, (b) axial–shear (A–S) coupling, (c) axial–rotation (A–R) coupling, (d) axial–bending (A–B) coupling, (e) shear–rotation (S–R) coupling, (f) shear–bending (S–B) coupling, (g) bending–rotation (B–R) coupling, and (h) bending–bending (B–B) coupling.*

We summarize the eight different coupling effects with the corresponding terms of **q** in Table 1. Similar to Poisson's ratio, we can define the coupling coefficients as the ratio of the two related strains, where $c_{ij}^{-1}$, $b_{ij}^{-1}$, and $d_{ij}^{-1}$ represent the elements of $\mathbf{q}^{-1}$. Note that one coupling effect can have multiple coupling coefficients depending on the direction of input strains.

*Table 1. 2D coupling effects and corresponding terms in the decoupled micropolar elasticity tensor*

| Names of mechanical coupling | Relevant elements in **q** | Coupling coefficients |
|---|---|---|
| Poisson's effect (A–A coupling) | $c_{12}$ | $c_{12}^{-1}/c_{11}^{-1}$ |
| Axial–shear (A–S) coupling | $c_{13}$, $c_{23}$ | $c_{13}^{-1}/c_{11}^{-1}$, $c_{23}^{-1}/c_{22}^{-1}$ |
| Axial–rotation (A–R) coupling | $c_{14}$, $c_{24}$ | $c_{14}^{-1}/c_{11}^{-1}$, $c_{24}^{-1}/c_{22}^{-1}$ |
| Axial–bending (A–B) coupling | $b_{11}$, $b_{12}$, $b_{21}$, $b_{22}$ | $b_{11}^{-1}/c_{11}^{-1}$, $b_{12}^{-1}/c_{11}^{-1}$, $b_{21}^{-1}/c_{22}^{-1}$, $b_{22}^{-1}/c_{22}^{-1}$ |
| Shear–rotation (S–R) coupling | $c_{34}$ | $c_{34}^{-1}/c_{33}^{-1}$ |
| Shear–bending (S–B) coupling | $b_{31}$, $b_{32}$ | $b_{31}^{-1}/c_{33}^{-1}$, $b_{32}^{-1}/c_{33}^{-1}$ |
| Bending–rotation (B–R) coupling | $b_{41}$, $b_{42}$ | $b_{41}^{-1}/c_{44}^{-1}$, $b_{42}^{-1}/c_{44}^{-1}$ |
| Bending–bending (B–B) coupling | $d_{12}$ | $d_{12}^{-1}/d_{11}^{-1}$ |

Notably, the shear strain $E_{12}$ and the local rotation $A_{12}$ are invariant because they are independent of the coordinate transformation [38]. However, the axial strain and the bending curvature are variant. The coupling by two invariant strains produces an invariant, i.e., an S–R coupling coefficient in Table 1. The couplings by invariant and variant strains generate a variant, i.e., A–S, A–R, S–B, and B–R couplings. The couplings by two variant strains produce a more complicated scenario: the Poisson's effect (A–A coupling) is variant in anisotropic materials; however, it becomes invariant in isotropic materials [35]. The bending–bending and axial–bending couplings are more complicated, as discussed in Section 4.

## 3. Symmetry operations of 2D lattices

All 2D lattice geometries can be classified into 10 point groups depending on their symmetry property [29,39]. There are seven symmetry operations in a 2D plane: the identity transformation (**1**), 2-fold rotation ($\mathbf{R_2}$), 3-fold rotation ($\mathbf{R_3}$), 4-fold rotation ($\mathbf{R_4}$), 6-fold rotation ($\mathbf{R_6}$), and mirror with respect to the $x_1$-axis ($\mathbf{m_1}$) and $x_2$ axis ($\mathbf{m_2}$). Every symmetry operation corresponds to a coordinate transformation by the direction cosine $l_{ij}(=\mathbf{L})$ between unbarred ($\hat{\mathbf{e}}_j$) and barred ($\hat{\bar{\mathbf{e}}}_i$) coordinate bases:

$$\hat{\bar{\mathbf{e}}}_i = l_{ij}\hat{\mathbf{e}}_j, \tag{8}$$

where $l_{ij} = \begin{bmatrix} 1 & 0 \\ 0 & 1 \end{bmatrix}$ for an identity transformation, $l_{ij} = \begin{bmatrix} \cos\theta & \sin\theta \\ -\sin\theta & \cos\theta \end{bmatrix}$ for a rotational transformation ($R_2$, $R_3$, $R_4$, and $R_6$ correspond to $\theta = \pi, \frac{2\pi}{3}, \frac{\pi}{2},$ and $\frac{\pi}{3}$, respectively), and $l_{ij} = \begin{bmatrix} 1 & 0 \\ 0 & -1 \end{bmatrix}$ and $l_{ij} = \begin{bmatrix} -1 & 0 \\ 0 & 1 \end{bmatrix}$ are the mirror transformations with respect to the $x_1$- and $x_2$ axes, respectively.

Using the transformation law of second-order tensors [31],

$$\begin{bmatrix} \bar{\sigma}_{11} \\ \bar{\sigma}_{22} \\ \bar{\sigma}_{12} \\ \bar{\sigma}_{21} \end{bmatrix} = |l| \begin{bmatrix} l_{11}^2 & l_{12}^2 & l_{11}l_{12} & l_{11}l_{12} \\ l_{21}^2 & l_{22}^2 & l_{21}l_{22} & l_{21}l_{22} \\ l_{11}l_{21} & l_{12}l_{22} & l_{11}l_{22} & l_{12}l_{21} \\ l_{11}l_{21} & l_{12}l_{22} & l_{12}l_{21} & l_{11}l_{22} \end{bmatrix} \begin{bmatrix} \sigma_{11} \\ \sigma_{22} \\ \sigma_{12} \\ \sigma_{21} \end{bmatrix}, \tag{9}$$

where $|l|$ is the determinant of $l_{ij}$. Note that $|l| = 1$ for a rotational transformation and $|l| = -1$ for a mirror transformation because the mirror transformation changes the handedness of the axial coordinate.

However, the couple stress in 2D is a first-order tensor; thus,

$$\begin{bmatrix} \bar{m}_{13} \\ \bar{m}_{23} \end{bmatrix} = |l| \begin{bmatrix} l_{11} & l_{12} \\ l_{21} & l_{22} \end{bmatrix} \begin{bmatrix} m_{13} \\ m_{23} \end{bmatrix}. \tag{10}$$

The transformation matrix for the generalized stress vector **σ** in Equation (2) can be assembled from Equations (9)–(10):

$$\mathbf{R} = |l| \begin{bmatrix} l_{11}^2 & l_{12}^2 & l_{11}l_{12} & l_{11}l_{12} & 0 & 0 \\ l_{21}^2 & l_{22}^2 & l_{21}l_{22} & l_{21}l_{22} & 0 & 0 \\ l_{11}l_{21} & l_{12}l_{22} & l_{11}l_{22} & l_{12}l_{21} & 0 & 0 \\ l_{11}l_{21} & l_{12}l_{22} & l_{12}l_{21} & l_{11}l_{22} & 0 & 0 \\ 0 & 0 & 0 & 0 & l_{11} & l_{12} \\ 0 & 0 & 0 & 0 & l_{21} & l_{22} \end{bmatrix}. \tag{11}$$

Thus, the transformed stress in a vector form is given by

$$\bar{\boldsymbol{\sigma}} = \mathbf{R} \cdot \boldsymbol{\sigma}. \tag{12}$$

According to [31], the transformed strain is given by

$$\bar{\boldsymbol{\varepsilon}} = (\mathbf{R}^T)^{-1} \cdot \boldsymbol{\varepsilon}. \tag{13}$$

Notably, **R** is an orthogonal transformation: $\mathbf{R}^{-T} = \mathbf{R}$.

The constitutive equation in the transformed coordinate is

$$\bar{\boldsymbol{\sigma}} = \bar{\mathbf{Q}} \cdot \bar{\boldsymbol{\varepsilon}}. \tag{14}$$

Substituting Equations (12) and (13) into Equation (14) gives:

$$\mathbf{R} \cdot \boldsymbol{\sigma} = \bar{\mathbf{Q}} \cdot \mathbf{R}^{-T} \cdot \boldsymbol{\varepsilon} \tag{15}$$

i.e.,

$$\boldsymbol{\sigma} = \mathbf{R}^{-1} \cdot \bar{\mathbf{Q}} \cdot \mathbf{R}^{-T} \cdot \boldsymbol{\varepsilon}. \tag{16}$$

Note the elasticity tensors in the original and transformed coordinates are given by

$$\mathbf{Q} = \mathbf{R}^{-1} \cdot \bar{\mathbf{Q}} \cdot \mathbf{R}^{-T} \tag{17}$$

$$\bar{\mathbf{Q}} = \mathbf{R} \cdot \mathbf{Q} \cdot \mathbf{R}^T. \tag{18}$$

According to Neumann's principle [31], if the geometry is invariant under such transformation, the elasticity tensor should also be invariant, i.e.,

$$\bar{\mathbf{Q}} = \mathbf{Q}. \tag{19}$$

Solving Equation (19) by comparing the elasticity tensor before and after symmetry operation, one can identify the nonzero components of **Q**. Table 2 summarizes the nonzero terms of **Q** for different symmetry operations, showing that each symmetry operation has different sets of nonzero terms.

*Table 2. Micropolar elasticity tensor $Q$ for varying symmetry operations*

| Symmetry operation | Elasticity tensor | No. of independent components |
|---|---|---|
| 1 | $\begin{bmatrix} C_{11} & C_{12} & C_{13} & C_{14} & B_{11} & B_{12} \\ & C_{22} & C_{23} & C_{24} & B_{21} & B_{22} \\ & & C_{33} & C_{34} & B_{31} & B_{32} \\ & & & C_{44} & B_{41} & B_{42} \\ & sym & & & D_{11} & D_{12} \\ & & & & & D_{22} \end{bmatrix}$ | 21 |
| $R_2$ | $\begin{bmatrix} C_{11} & C_{12} & C_{13} & C_{14} & 0 & 0 \\ & C_{22} & C_{23} & C_{24} & 0 & 0 \\ & & C_{33} & C_{34} & 0 & 0 \\ & & & C_{44} & 0 & 0 \\ & sym & & & D_{11} & D_{12} \\ & & & & & D_{22} \end{bmatrix}$ | 13 |
| $R_3$ | $\begin{bmatrix} C_{11} & C_{12} & C_{13} & -C_{13} & B_{11} & B_{12} \\ & C_{11} & C_{13} & -C_{13} & -B_{11} & -B_{12} \\ & & C_{33} & C_{11}-C_{12}-C_{33} & B_{12} & -B_{11} \\ & & & C_{33} & B_{12} & -B_{11} \\ & sym & & & D_{11} & 0 \\ & & & & & D_{11} \end{bmatrix}$ | 7 |
| $R_4$ | $\begin{bmatrix} C_{11} & C_{12} & C_{13} & C_{14} & 0 & 0 \\ & C_{11} & -C_{14} & -C_{13} & 0 & 0 \\ & & C_{33} & C_{34} & 0 & 0 \\ & & & C_{33} & 0 & 0 \\ & sym & & & D_{11} & 0 \\ & & & & & D_{11} \end{bmatrix}$ | 7 |
| $R_6$ | $\begin{bmatrix} C_{11} & C_{12} & C_{13} & -C_{13} & 0 & 0 \\ & C_{11} & C_{13} & -C_{13} & 0 & 0 \\ & & C_{33} & C_{11}-C_{12}-C_{33} & 0 & 0 \\ & & & C_{33} & 0 & 0 \\ & sym & & & D_{11} & 0 \\ & & & & & D_{11} \end{bmatrix}$ | 5 |
| $m_1$ | $\begin{bmatrix} C_{11} & C_{12} & 0 & 0 & B_{11} & 0 \\ & C_{22} & 0 & 0 & B_{21} & 0 \\ & & C_{33} & C_{34} & 0 & B_{32} \\ & & & C_{44} & 0 & B_{42} \\ & sym & & & D_{11} & 0 \\ & & & & & D_{22} \end{bmatrix}$ | 12 |

| Symmetry operation | | No. |
|---|---|---|
| $m_2$ | $\begin{bmatrix} C_{11} & C_{12} & 0 & 0 & 0 & B_{12} \\ & C_{22} & 0 & 0 & 0 & B_{22} \\ & & C_{33} & C_{34} & B_{31} & 0 \\ & & & C_{44} & B_{41} & 0 \\ & \text{sym} & & & D_{11} & 0 \\ & & & & & D_{22} \end{bmatrix}$ | 12 |

The components of $\begin{bmatrix} B_{11} & B_{12} \\ B_{21} & B_{22} \end{bmatrix}$ are 0 if and only if the geometry has an $R_2$ symmetry, which is consistent with a previous study [40]. Note that $R_2$ includes $R_4$ and $R_6$. Also note that the 'centrosymmetric' used in the literature [18,41] is equivalent to $R_2$ for 2D structures.

Applying Equation (7) to Table 2, we can obtain the decoupled micropolar elasticity tensor **q** for each symmetry operation, showing that each symmetry operation provides different coupling effects in the non-diagonal elements of the decoupled elasticity tensor, as shown in Table 3.

*Table 3. Decoupled micropolar elasticity tensor **q** for varying symmetry operations*

| Symmetry operation | Decoupled elasticity tensor | No. of independent variables |
|---|---|---|
| 1 | $\begin{bmatrix} c_{11} & c_{12} & c_{13} & c_{14} & b_{11} & b_{12} \\ & c_{22} & c_{23} & c_{24} & b_{21} & b_{22} \\ & & c_{33} & c_{34} & b_{31} & b_{32} \\ & & & c_{44} & b_{41} & b_{42} \\ & \text{sym} & & & d_{11} & d_{12} \\ & & & & & d_{22} \end{bmatrix}$ | 21 |
| $R_2$ | $\begin{bmatrix} c_{11} & c_{12} & c_{13} & c_{14} & 0 & 0 \\ & c_{22} & c_{23} & c_{24} & 0 & 0 \\ & & c_{33} & c_{34} & 0 & 0 \\ & & & c_{44} & 0 & 0 \\ & \text{sym} & & & d_{11} & d_{12} \\ & & & & & d_{22} \end{bmatrix}$ | 13 |
| $R_3$ | $\begin{bmatrix} c_{11} & c_{12} & 0 & c_{14} & b_{11} & b_{12} \\ & c_{11} & 0 & c_{14} & -b_{11} & -b_{12} \\ & & \frac{c_{11}-c_{12}}{2} & 0 & b_{12} & -b_{11} \\ & & & c_{44} & 0 & 0 \\ & \text{sym} & & & d_{11} & 0 \\ & & & & & d_{11} \end{bmatrix}$ | 7 |
| $R_4$ | $\begin{bmatrix} c_{11} & c_{12} & c_{13} & c_{14} & 0 & 0 \\ & c_{11} & -c_{13} & c_{14} & 0 & 0 \\ & & c_{33} & 0 & 0 & 0 \\ & & & c_{44} & 0 & 0 \\ & \text{sym} & & & d_{11} & 0 \\ & & & & & d_{11} \end{bmatrix}$ | 7 |
| $R_6$ | $\begin{bmatrix} c_{11} & c_{12} & 0 & c_{14} & 0 & 0 \\ & c_{11} & 0 & c_{14} & 0 & 0 \\ & & \frac{c_{11}-c_{12}}{2} & 0 & 0 & 0 \\ & & & c_{44} & 0 & 0 \\ & sym & & & d_{11} & 0 \\ & & & & & d_{11} \end{bmatrix}$ | 5 |
| $m_1$ | $\begin{bmatrix} c_{11} & c_{12} & 0 & 0 & b_{11} & 0 \\ & c_{22} & 0 & 0 & b_{21} & 0 \\ & & c_{33} & c_{34} & 0 & b_{32} \\ & & & c_{44} & 0 & b_{42} \\ & sym & & & d_{11} & 0 \\ & & & & & d_{22} \end{bmatrix}$ | 12 |
| $m_2$ | $\begin{bmatrix} c_{11} & c_{12} & 0 & 0 & 0 & b_{12} \\ & c_{22} & 0 & 0 & 0 & b_{22} \\ & & c_{33} & c_{34} & b_{31} & 0 \\ & & & c_{44} & b_{41} & 0 \\ & sym & & & d_{11} & 0 \\ & & & & & d_{22} \end{bmatrix}$ | 12 |

Apparently, adding symmetry leads to a reduction of mechanical coupling effects. The $R_2$ symmetry eliminates A–B, S–B, and B–R couplings. The $R_4$ symmetry further eliminates an S–R coupling. The $R_3$ symmetry ensures that there is no A–S coupling, S–R coupling, B–R coupling, or B–B coupling. Because $R_6$ is a combination of $R_2$ and $R_3$, it can eliminate all the coupling effects except Poisson's effect and an A–R coupling. The $m_1$ and $m_2$ symmetry operations eliminate A–S and A–R couplings. Notably, no symmetry operation can eliminate Poisson's effect.

Yuan et al. showed that structures with an $R_4$ symmetry such as tetra-chiral structures have an A–S coupling [15]. Liu et al. showed that a hexa-chiral structure with an $R_6$ symmetry has a bulk–rotation coupling [17], which is equivalent to an A–R coupling of this work. Cui et al. showed that the non-centrosymmetric square lattice has an A–B coupling [18]. However, these works [15,17,18] lack theoretical proof of the relationship between coupling effects and symmetry. In contrast, our present work reveals that symmetry is a necessary but not sufficient condition for the existence of A–S, A–R, A–B, and other coupling effects.

## 4. Mechanical coupling in the point groups

In general, a lattice structure can possess more than one symmetry operation; a lattice geometry can be classified into multiple point groups based on the symmetry operations to which it belongs [39]. The micropolar elasticity tensor of a point group must remain unchanged under all the symmetry operations of this point group. There are ten point groups in 2D lattice structures: 1, 2, $m$, $2mm$, 4, $4mm$, 3, $3m$, 6, and $6mm$. From Table 1, we can classify the micropolar elasticity tensor for every point group together with the corresponding symmetry operations, as shown in Table 4. Note that there are two types of elasticity tensors for the point groups $m$ and $3m$ depending on the mirror symmetry axes ($x_1$ and $x_2$), as indicated in Table 4. Thus, we have $12(=10+2)$ types of micropolar elasticity tensor **Q** compared with three types in the Cauchy elasticity tensor [39]. The point group 3 ensures isotropy in the Cauchy elasticity [42]. However, the point group $6mm$ ensures an isotropy in the micropolar elasticity because more parameters are used in the micropolar theory to describe higher-order material properties. We can also find that the A–B and B–B couplings are not invariant because these coupling effects disappear in the elasticity tensor of $6mm$; there are no $B_{ij}$ (or $b_{ij}$) or $D_{12}$ (or $d_{12}$) elements in $6mm$ in Tables 4 and 5.

*Table 4. Micropolar elasticity tensor $\mathbf{Q}$ of each point group*

| Point group | Symmetry operations | Elasticity tensor | No. of independent variables |
|---|---|---|---|
| 1 | 1 | $\begin{bmatrix} C_{11} & C_{12} & C_{13} & C_{14} & B_{11} & B_{12} \\ & C_{22} & C_{23} & C_{24} & B_{21} & B_{22} \\ & & C_{33} & C_{34} & B_{31} & B_{32} \\ & & & C_{44} & B_{41} & B_{42} \\ & sym & & & D_{11} & D_{12} \\ & & & & & D_{22} \end{bmatrix}$ | 21 |
| 2 | $R_2$ | $\begin{bmatrix} C_{11} & C_{12} & C_{13} & C_{14} & 0 & 0 \\ & C_{22} & C_{23} & C_{24} & 0 & 0 \\ & & C_{33} & C_{34} & 0 & 0 \\ & & & C_{44} & 0 & 0 \\ & sym & & & D_{11} & D_{12} \\ & & & & & D_{22} \end{bmatrix}$ | 13 |
| m | $m_1$ | $\begin{bmatrix} C_{11} & C_{12} & 0 & 0 & B_{11} & 0 \\ & C_{22} & 0 & 0 & B_{21} & 0 \\ & & C_{33} & C_{34} & 0 & B_{32} \\ & & & C_{44} & 0 & B_{42} \\ & sym & & & D_{11} & 0 \\ & & & & & D_{22} \end{bmatrix}$ | 12 |
| m | $m_2$ | $\begin{bmatrix} C_{11} & C_{12} & 0 & 0 & 0 & B_{12} \\ & C_{22} & 0 & 0 & 0 & B_{22} \\ & & C_{33} & C_{34} & B_{31} & 0 \\ & & & C_{44} & B_{41} & 0 \\ & sym & & & D_{11} & 0 \\ & & & & & D_{22} \end{bmatrix}$ | 12 |
| 2mm | $R_2, m_1, m_2$ | $\begin{bmatrix} C_{11} & C_{12} & 0 & 0 & 0 & 0 \\ & C_{22} & 0 & 0 & 0 & 0 \\ & & C_{33} & C_{34} & 0 & 0 \\ & & & C_{44} & 0 & 0 \\ & sym & & & D_{11} & 0 \\ & & & & & D_{22} \end{bmatrix}$ | 8 |
| 4 | $R_4$ | $\begin{bmatrix} C_{11} & C_{12} & C_{13} & C_{14} & 0 & 0 \\ & C_{11} & -C_{14} & -C_{13} & 0 & 0 \\ & & C_{33} & C_{34} & 0 & 0 \\ & & & C_{33} & 0 & 0 \\ & sym & & & D_{11} & 0 \\ & & & & & D_{11} \end{bmatrix}$ | 7 |
| 4mm | $R_4, m_1, m_2$ | $\begin{bmatrix} C_{11} & C_{12} & 0 & 0 & 0 & 0 \\ & C_{11} & 0 & 0 & 0 & 0 \\ & & C_{33} & C_{34} & 0 & 0 \\ & & & C_{33} & 0 & 0 \\ & sym & & & D_{11} & 0 \\ & & & & & D_{11} \end{bmatrix}$ | 5 |
| 3 | $R_3$ | $\begin{bmatrix} C_{11} & C_{12} & C_{13} & -C_{13} & B_{11} & B_{12} \\ & C_{11} & C_{13} & -C_{13} & -B_{11} & -B_{12} \\ & & C_{33} & C_{11}-C_{12}-C_{33} & B_{12} & -B_{11} \\ & & & C_{33} & B_{12} & -B_{11} \\ & sym & & & D_{11} & 0 \\ & & & & & D_{11} \end{bmatrix}$ | 7 |
| 3m | $R_3, m_1$ | $\begin{bmatrix} C_{11} & C_{12} & 0 & 0 & B_{11} & 0 \\ & C_{11} & 0 & 0 & -B_{11} & 0 \\ & & C_{33} & C_{11}-C_{12}-C_{33} & 0 & -B_{11} \\ & & & C_{33} & 0 & -B_{11} \\ & sym & & & D_{11} & 0 \\ & & & & & D_{11} \end{bmatrix}$ | 5 |
| 3m | $R_3, m_2$ | $\begin{bmatrix} C_{11} & C_{12} & 0 & 0 & 0 & B_{12} \\ & C_{11} & 0 & 0 & 0 & -B_{12} \\ & & C_{33} & C_{11}-C_{12}-C_{33} & B_{12} & 0 \\ & & & C_{33} & B_{12} & 0 \\ & sym & & & D_{11} & 0 \\ & & & & & D_{11} \end{bmatrix}$ | 5 |
| 6 | $R_6$ | $\begin{bmatrix} C_{11} & C_{12} & C_{13} & -C_{13} & 0 & 0 \\ & C_{11} & C_{13} & -C_{13} & 0 & 0 \\ & & C_{33} & C_{11}-C_{12}-C_{33} & 0 & 0 \\ & & & C_{33} & 0 & 0 \\ & sym & & & D_{11} & 0 \\ & & & & & D_{11} \end{bmatrix}$ | 5 |
| 6mm | $R_6, m_1, m_2$ | $\begin{bmatrix} C_{11} & C_{12} & 0 & 0 & 0 & 0 \\ & C_{11} & 0 & 0 & 0 & 0 \\ & & C_{33} & C_{11}-C_{12}-C_{33} & 0 & 0 \\ & & & C_{33} & 0 & 0 \\ & sym & & & D_{11} & 0 \\ & & & & & D_{11} \end{bmatrix}$ | 4 |

We can compare the elasticity tensor **Q** with the ones in the literature for different lattice structures, such as a straight rectangle (point group $2mm$) [28], a chiral square (point group 4) [40], a straight square (point group $4mm$) [28], a chiral triangle (point group 6) [17], a straight triangle (point group $6mm$) [28], and a non-centrosymmetric square (point group $m$) [18]. The corresponding lattices are shown in Figure 4. The elasticity tensors of all the structures in the literature match well with the ones in Table 4. Note that Liu [17] and Spadoni [9] reported different results for a chiral triangular structure, with Liu's model being more accurate and compatible with our theory.

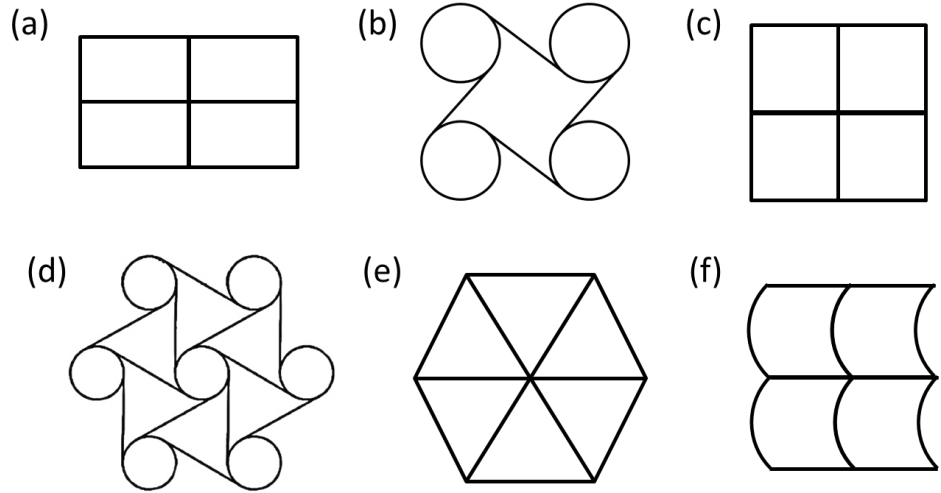

*Figure 4. Geometries of (a) straight rectangle (point group $2mm$) [28], (b) chiral square (point group 4) [40], (c) straight square (point group $4mm$) [28], (d) chiral triangle (point group 6) [17], (e) straight triangle (point group $6mm$) [28], and (f) non-centrosymmetric square (point group $m$) [18]*

Similarly, we can obtain the relationship between the point groups and the decoupled micropolar elasticity tensor **q**, given in Table 5. Notably, each point group can classify the non-diagonal elements of **q** in Table 5, providing a design guideline of mechanical coupling with a selection of lattice geometries.

*Table 5. Decoupled micropolar elasticity tensor **q** for each point group*

| Point group | Symmetry operations | Decoupled elasticity tensor | No. of independent variables |
|---|---|---|---|
| 1 | 1 | $\begin{bmatrix} c_{11} & c_{12} & c_{13} & c_{14} & b_{11} & b_{12} \\ & c_{22} & c_{23} & c_{24} & b_{21} & b_{22} \\ & & c_{33} & c_{34} & b_{31} & b_{32} \\ & & & c_{44} & b_{41} & b_{42} \\ & sym & & & d_{11} & d_{12} \\ & & & & & d_{22} \end{bmatrix}$ | 21 |
| 2 | $R_2$ | $\begin{bmatrix} c_{11} & c_{12} & c_{13} & c_{14} & 0 & 0 \\ & c_{22} & c_{23} & c_{24} & 0 & 0 \\ & & c_{33} & c_{34} & 0 & 0 \\ & & & c_{44} & 0 & 0 \\ & sym & & & d_{11} & d_{12} \\ & & & & & d_{22} \end{bmatrix}$ | 13 |
| m | $m_1$ | $\begin{bmatrix} c_{11} & c_{12} & 0 & 0 & b_{11} & 0 \\ & c_{22} & 0 & 0 & b_{21} & 0 \\ & & c_{33} & c_{34} & 0 & b_{32} \\ & & & c_{44} & 0 & b_{42} \\ & sym & & & d_{11} & 0 \\ & & & & & d_{22} \end{bmatrix}$ | 12 |
| | $m_2$ | $\begin{bmatrix} c_{11} & c_{12} & 0 & 0 & 0 & b_{12} \\ & c_{22} & 0 & 0 & 0 & b_{22} \\ & & c_{33} & c_{34} & b_{31} & 0 \\ & & & c_{44} & b_{41} & 0 \\ & sym & & & d_{11} & 0 \\ & & & & & d_{22} \end{bmatrix}$ | 12 |
| 2mm | $R_2, m_1, m_2$ | $\begin{bmatrix} c_{11} & c_{12} & 0 & 0 & 0 & 0 \\ & c_{22} & 0 & 0 & 0 & 0 \\ & & c_{33} & c_{34} & 0 & 0 \\ & & & c_{44} & 0 & 0 \\ & sym & & & d_{11} & 0 \\ & & & & & d_{22} \end{bmatrix}$ | 8 |
| 4 | $R_4$ | $\begin{bmatrix} c_{11} & c_{12} & c_{13} & c_{14} & 0 & 0 \\ & c_{11} & -c_{13} & c_{14} & 0 & 0 \\ & & c_{33} & 0 & 0 & 0 \\ & & & c_{44} & 0 & 0 \\ & sym & & & d_{11} & 0 \\ & & & & & d_{11} \end{bmatrix}$ | 7 |
| 4mm | $R_4, m_1, m_2$ | $\begin{bmatrix} c_{11} & c_{12} & 0 & 0 & 0 & 0 \\ & c_{11} & 0 & 0 & 0 & 0 \\ & & c_{33} & 0 & 0 & 0 \\ & & & c_{44} & 0 & 0 \\ & sym & & & d_{11} & 0 \\ & & & & & d_{11} \end{bmatrix}$ | 5 |
| 3 | $R_3$ | $\begin{bmatrix} c_{11} & c_{12} & 0 & c_{14} & b_{11} & b_{12} \\ & c_{11} & 0 & c_{14} & -b_{11} & -b_{12} \\ & & \frac{c_{11}-c_{12}}{2} & 0 & b_{12} & -b_{11} \\ & & & c_{44} & 0 & 0 \\ & sym & & & d_{11} & 0 \\ & & & & & d_{11} \end{bmatrix}$ | 7 |
| 3m | $R_3, m_1$ | $\begin{bmatrix} c_{11} & c_{12} & 0 & 0 & b_{11} & 0 \\ & c_{11} & 0 & 0 & -b_{11} & 0 \\ & & \frac{c_{11}-c_{12}}{2} & 0 & 0 & -b_{11} \\ & & & c_{44} & 0 & 0 \\ & sym & & & d_{11} & 0 \\ & & & & & d_{11} \end{bmatrix}$ | 5 |
| | $R_3, m_2$ | $\begin{bmatrix} c_{11} & c_{12} & 0 & 0 & 0 & b_{12} \\ & c_{11} & 0 & 0 & 0 & -b_{12} \\ & & \frac{c_{11}-c_{12}}{2} & 0 & b_{12} & 0 \\ & & & c_{44} & 0 & 0 \\ & sym & & & d_{11} & 0 \\ & & & & & d_{11} \end{bmatrix}$ | 5 |
| 6 | $R_6$ | $\begin{bmatrix} c_{11} & c_{12} & 0 & c_{14} & 0 & 0 \\ & c_{11} & 0 & c_{14} & 0 & 0 \\ & & \frac{c_{11}-c_{12}}{2} & 0 & 0 & 0 \\ & & & c_{44} & 0 & 0 \\ & sym & & & d_{11} & 0 \\ & & & & & d_{11} \end{bmatrix}$ | 5 |
| 6mm | $R_6, m_1, m_2$ | $\begin{bmatrix} c_{11} & c_{12} & 0 & 0 & 0 & 0 \\ & c_{11} & 0 & 0 & 0 & 0 \\ & & \frac{c_{11}-c_{12}}{2} & 0 & 0 & 0 \\ & & & c_{44} & 0 & 0 \\ & sym & & & d_{11} & 0 \\ & & & & & d_{11} \end{bmatrix}$ | 4 |

## 5. Mechanical coupling with geometric symmetry

Table 6 summarizes the relationship between the point groups and mechanical couplings, together with representative lattices. Among the eight coupling effects in Figure 2, the A–A coupling (Poisson's effect) belongs to all point groups, whereas the others gradually disappear as the degree of symmetry increases. Some of our findings are consistent with previous works; i.e., a tetra-chiral structure (point group 4) shows an A–S coupling [15,36], a triangular chiral lattice (point group 6) shows an A–R coupling [17], and a non-centrosymmetric square lattice (point group $m$) shows an A–B coupling [18]. Neumann's principle can find zero terms only in the elasticity tensor without considering geometric connection, meaning the nonzero terms could be zero for a specific geometry. Therefore, it may be fair to say that the point group is a necessary but not sufficient condition of the coupling effects. For example, we can ensure that the square lattice whose point group is $4mm$ does not have an A–B coupling, as shown in Table 6, but cannot ensure that all the geometries belonging to $4mm$ have an A–A coupling. In fact, Poisson's ratio of a straight square lattice loaded in the principal direction is 0, showing no A–A coupling effect [43].

*Table 6. Relationships between mechanical coupling effects and point groups*

| Point group | Symmetry operation | Representative lattice | Mechanical coupling |
|---|---|---|---|
| 1 | 1 | 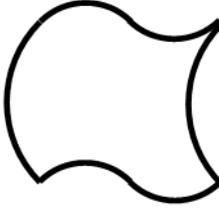 | A–A, A–S, A–R, A–B, S–R, S–B, B–R, B–B |
| 2 | $R_2$ | 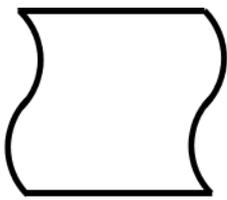 | A–A, A–S, A–R, S–R, B–B |
| $m$ | $m_1$ or $m_2$ | 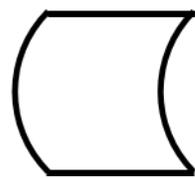 | A–A, A–B, S–R, S–B, B–R |
| $2mm$ | $R_2, m_1, m_2$ | 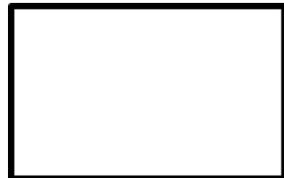 | A–A, S–R |
| 4 | $R_4$ | 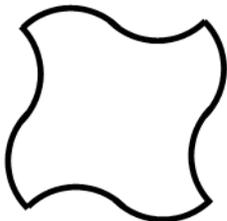 | A–A, A–S, A–R |
| $4mm$ | $R_4, m_1, m_2$ | 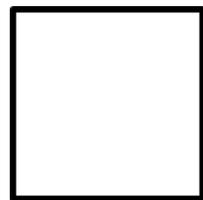 | A–A |
| 3 | $R_3$ | 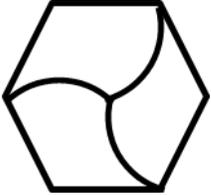 | A–A, A–R, A–B, S–B |
| 3m | $R_3, m_1$ or $R_3, m_2$ | 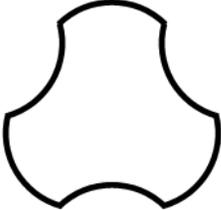 | A–A, A–B, S–B |
| 6 | $R_6$ | 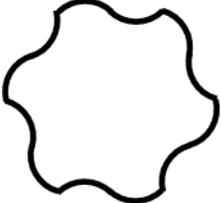 | A–A, A–R |
| 6mm | $R_6, m_1, m_2$ | 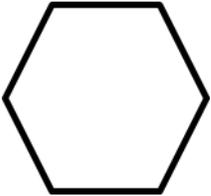 | A–A |

Table 7 summarizes the relationship between the coupling effects and point groups. Notably, all point groups have an axial–axial coupling effect. However, a chance of some coupling effects is given to a specific point group; e.g., only point groups 1 and $m$ can have a bending–rotation coupling. Only point groups 1 and 2 may have a bending–bending coupling. Note that the point groups are a necessary but not sufficient condition for the corresponding coupling effects in Table 7.

*Table 7. 2D coupling effects and the corresponding point groups*

| Coupling effect | Existence in the point groups |
|---|---|
| Axial–axial coupling (Poisson's effect) | all |
| Axial–shear coupling | 1, 2, 4 |
| Axial–rotation coupling | 1, 2, 3, 4, 6 |
| Axial–bending coupling | 1, 3, $m$, $3m$ |
| Shear–rotation coupling | 1, 2, $m$, $2mm$ |
| Shear–bending coupling | 1, 3, $m$, $3m$ |
| Bending–rotation coupling | 1, $m$ |
| Bending–bending coupling | 1, 2 |

## 6. Verification with numerical simulations

We design several lattice structures to demonstrate mechanical coupling effects, including the A–R, A–B, S–R, S–B, B–R, and B–B couplings. As shown in Table 7, we may find an axial–bending coupling effect in the point groups 1, $m$, 3, and $3m$. To avoid the effect of other coupling effects in point groups 1, 3, and $3m$, a structure of point group $m$ is constructed, as shown in Figure 5a. Nodal forces are applied on the left and right boundary nodes while maintaining the center node fixed. For comparison, we investigate the deformation of a straight square lattice of the point group $4mm$ for the same loading, as indicated in Figure 5b. Figure 5c plots the vertical displacements of the joints on the horizontal centerlines. Figure 5c shows the bending deformation of the horizontal centerline of a curved square lattice in the point group $m$ and no bending deformation of the straight square lattice belonging to the point group $4mm$.

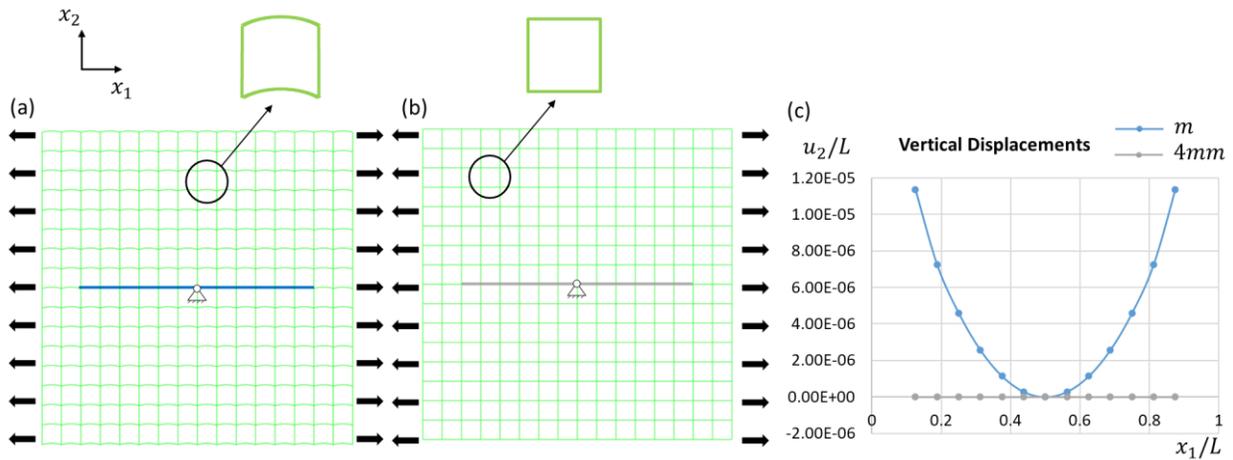

Figure 5. A tensile loading (normalized strain in the horizontal direction $E_{11} = 0.1\%$) applied to (a) a curved square lattice of point group $m$ and (b) a straight square lattice of point group $4mm$. (c) The vertical displacement of the nodes on the horizontal centerlines; the curved square lattice of the point group $m$ shows an A–B coupling.

A shear–rotation coupling means a nonzero shear strain $E_{12}$ producing a rotation $A_{12}$ in Equation (4), which exists in a rectangular lattice of the point group $2mm$ but not in a square lattice of the point group $4mm$. We apply a pure shear loading on the rectangular and square lattices, as shown in Figure 6a and Figure 6b. The rotational displacements $\varphi$ of the nodes on the horizontal centerlines are plotted in Figure 6c, demonstrating nonzero values for the rectangular lattice of $2mm$.

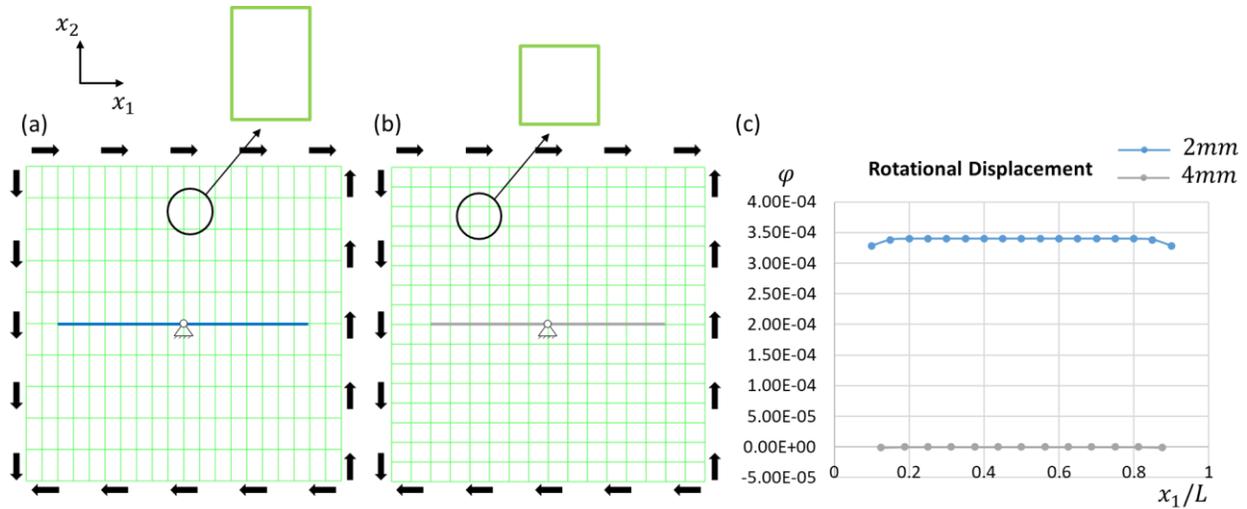

Figure 6. A pure shear loading ($E_{12} = 0.001$) applied to (a) a rectangular lattice of point group $2mm$ and (b) a square lattice of point group $4mm$. (c) Rotational displacements $\varphi$ of the nodes on the horizontal centerlines; the rectangular lattice of point group $2mm$ shows nonzero $\varphi$.

As shown in Table 7, S–B coupling can exist in point groups $m$, 3, and $3m$. We construct a structure belonging to $m$, as shown in Figure 7a. Because an S–B coupling is equivalent to a B–S coupling, we demonstrate a B–S coupling for convenience of measure. We apply moments on the nodes of the top and bottom boundaries. The vertical displacements of the nodes on the horizontal centerlines are plotted in Figure 7c, which shows the existence of shear deformation of the lattice $m$.

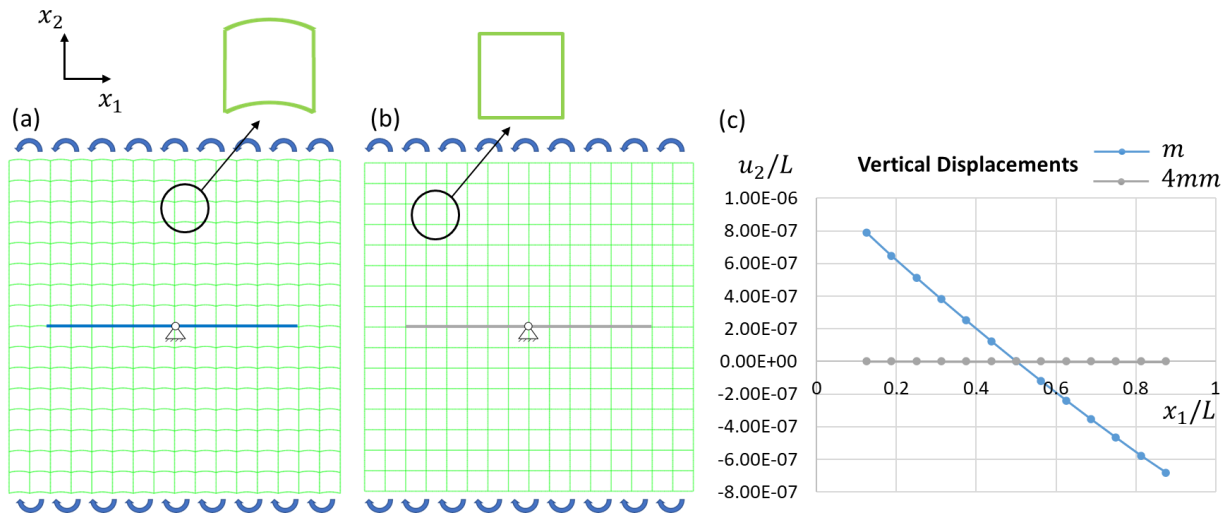

Figure 7. A bending loading ($\kappa_{23} = 0.1$) applied to (a) a curved square lattice of the point group m; (b) a straight square lattice of the point group $4mm$; (c) the vertical displacements on the nodes of the horizontal centerlines; the square lattice of the point group $m$ shows a B–S coupling.

Table 6 shows that a square lattice of the point group $m$ can have the bending–rotation (B–R) coupling. We apply moments on the top and bottom boundaries of square lattices of the point groups $m$ and $4mm$ as shown in Figures 8a and 8b, respectively, while fixing the center node. Figure 8c plots the rotational displacements $\varphi$ of the nodes on the horizontal centerline.

A straight square lattice of $4mm$ shows zero rotational displacements on the nodes of the horizontal centerline. In contrast, the curved square lattice of the point group $m$ shows a nonzero constant $\varphi$, verifying our theory.

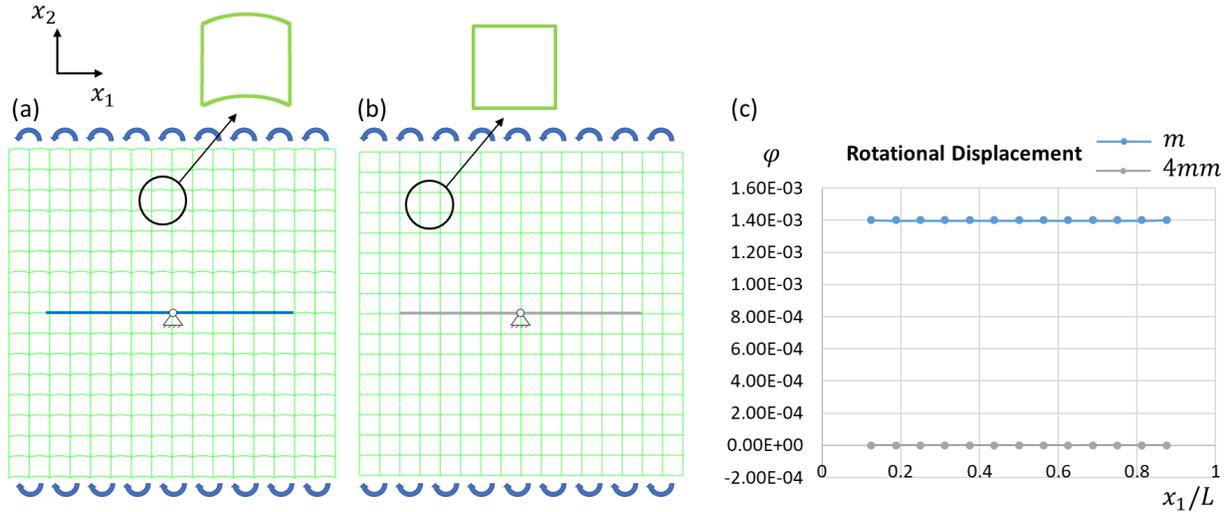

*Figure 8. A bending loading ($\kappa_{23} = 0.1$) applied to (a) a curved square lattice of point group $m$ and (b) a straight square lattice of point group $4mm$; (c) the rotational displacement of nodes on the horizontal centerlines; the curved square lattice of the point group $m$ demonstrates a B–R coupling.*

According to Table 6, a bending–bending (B–B) coupling exists in point groups 1 and 2. We select a curved lattice of point group 2 for a numerical test, as shown in Figure 9a. We apply moments on the top and bottom boundaries of the lattices. Figure 9c shows the horizontal displacements on the nodes of the vertical centerline. Apparently, there is a bending deformation in the horizontal direction for the curved rectangular lattice in point group 2, demonstrating a B–B coupling. However, the straight square lattice in point group $4mm$ does not show a B–B coupling.

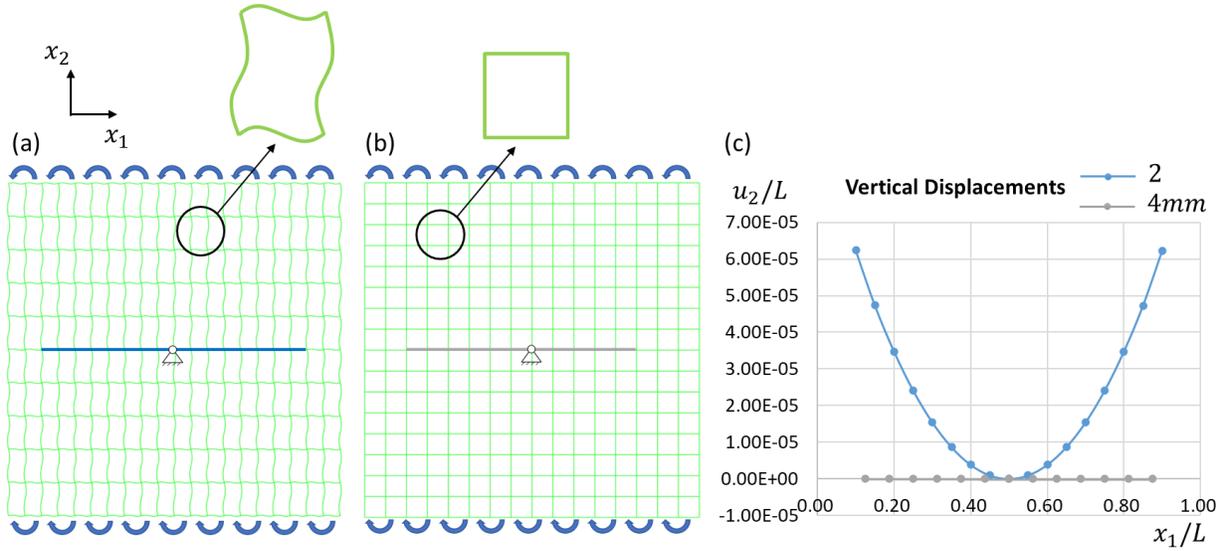

*Figure 9. A bending loading ($\kappa_{23} = 0.1$) applied to (a) a curved rectangular lattice of point group 2 and (b) a straight square lattice of point group $4mm$. (c) Horizontal displacement of nodes on the vertical centerlines; the curved rectangular lattice of point group 2 demonstrates a B–B coupling.*

There are many ways to generate mechanical couplings by breaking symmetry. Replacing straight ligaments of lattices with curved ones is one way to produce the coupling effects [15,18]. Alternative methods include assigning different materials to straight lattices [14], nonuniform cross-sections to the straight ligaments [44–46], and a combination of rigid rings and flexible struts [17,40,47]. Essentially, any methods that break the symmetry of the lattices can be used to design mechanical coupling effects.

## 7. Discussion

Due to no clear understanding of i) mechanical coupling with elasticity and ii) correlation between mechanical coupling and geometric symmetry, we secure the solution from the elasticity tensor and symmetry operation. We obtain the non-zero diagonal terms in the elasticity tensor by decoupling the shear deformation and nodal rotation in the micropolar elasticity tensor, revealing eight mechanical couplings in 2D lattices. Applying the symmetry operation to the decomposed micropolar elasticity tensor, we correlate the mechanical coupling with the point groups in 2D lattices. We also connect each mechanical coupling with the 10 point groups and seven symmetry operations in the planar lattice geometries.

This work focuses on the general guideline to identify mechanical couplings from the micropolar constitutive relations, limiting its scope to a qualitative study. However, one can extend our findings to quantitatively calculate mechanical coupling coefficients for a specific lattice geometry, e.g., a quantitative study of an A-S coupling coefficient of a tetra-chiral lattice [15]. Even though we can verify the mechanical couplings with FE simulations, some couplings are challenged to validate with a conventional mechanical testing method, e.g., B-R and B-B couplings of (g) and (h) in Figure 3. To rotate struts, one may need non-contact actuation, such as magnetic control.

Our methodology and findings are significant because they can provide a mechanics-inspired design guideline on the mechanical couplings integrated with geometric symmetry. Four out of the eight mechanical couplings are the first discovery in the history of mechanics: S-R, S-B, B-R, and B-B couplings. The correlation of the mechanical couplings with the point groups in this work can strengthen the design of mechanical metamaterials with potential applications in various areas, including sensors, actuators, soft robots, and active metamaterials.

## 8. Conclusion

We introduce a generalized methodology to uncover all mechanical couplings in 2D lattice geometries by obtaining the decoupled micropolar elasticity tensor. We also correlate the mechanical couplings with the point groups of 2D lattices by applying the symmetry operation to the decoupled micropolar elasticity tensor. The decoupled micropolar constitutive equation reveals eight mechanical coupling effects in planar solids, four of which are discovered for the first time in the mechanics community. The symmetry operation of the decoupled micropolar elasticity tensor shows the correlation of the mechanical coupling with the point groups. The correlation of the mechanical couplings with the point groups can strengthen the design of mechanical metamaterials with potential applications in areas including sensors, actuators, soft robots, and active metamaterials for elastic/acoustic wave guidance and thermal management. Our mechanics-inspired approach to discover the mechanical-coupling effects provides robust design guidelines to the metamaterial community, broadening the design space of mechanical metamaterials.

## 9. Acknowledgments

The authors acknowledge the support received from the Shanghai NSF (Award # 17ZR1414700) and the Research Incentive Program of Recruited Non-Chinese Foreign Faculty by Shanghai Jiao Tong University.

## 10. Author contributions

Z. Cui and J. Ju designed the research; Z. Cui designed and analyzed the lattice and conducted the continuum modeling and programming; Z. Cui performed the FE simulations; Z. Cui wrote the original draft; J. Ju wrote the paper with review and editing.

## 11. Appendix

We calculate the micropolar elasticity tensor of the representative lattices, implementing a finite element discrete modeling of the micropolar continuum [18,48]. The deformed configurations of different materials upon different loading conditions are presented in Figure A1, illustrating the A–A, A–S, A–R, A–B, S–R, S–B, B–R, and B–B coupling.

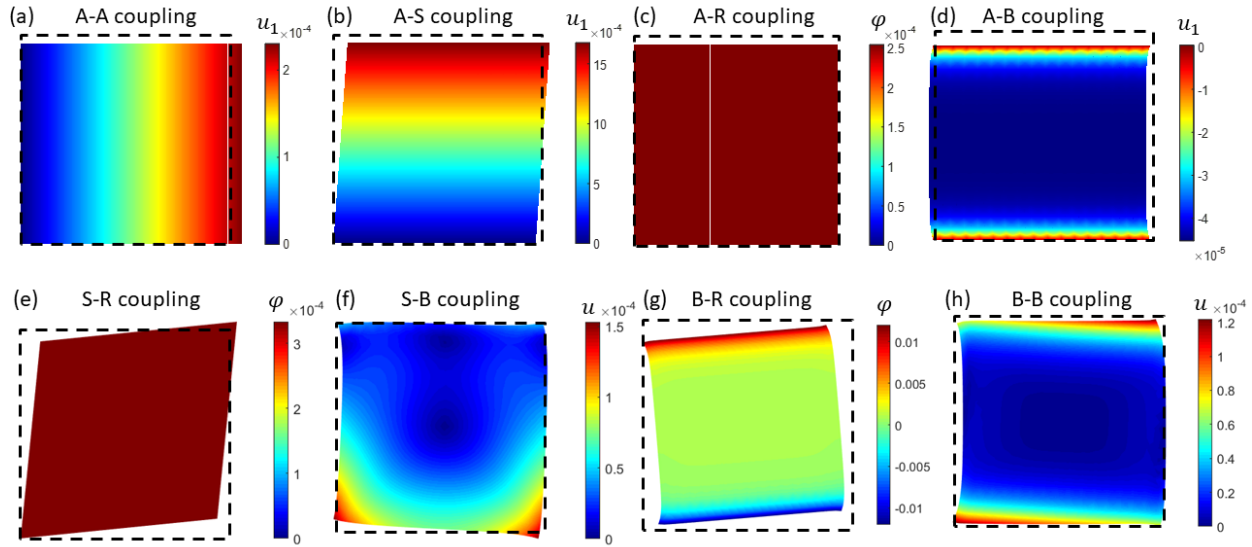

*Figure A1. Continuum modeling of (a) A–A coupling, (b) A–S coupling, (c) A–R coupling, (d) A–B coupling, (e) S–R coupling, (f) S–B coupling, (g) B–R coupling, and (h) B–B coupling.*